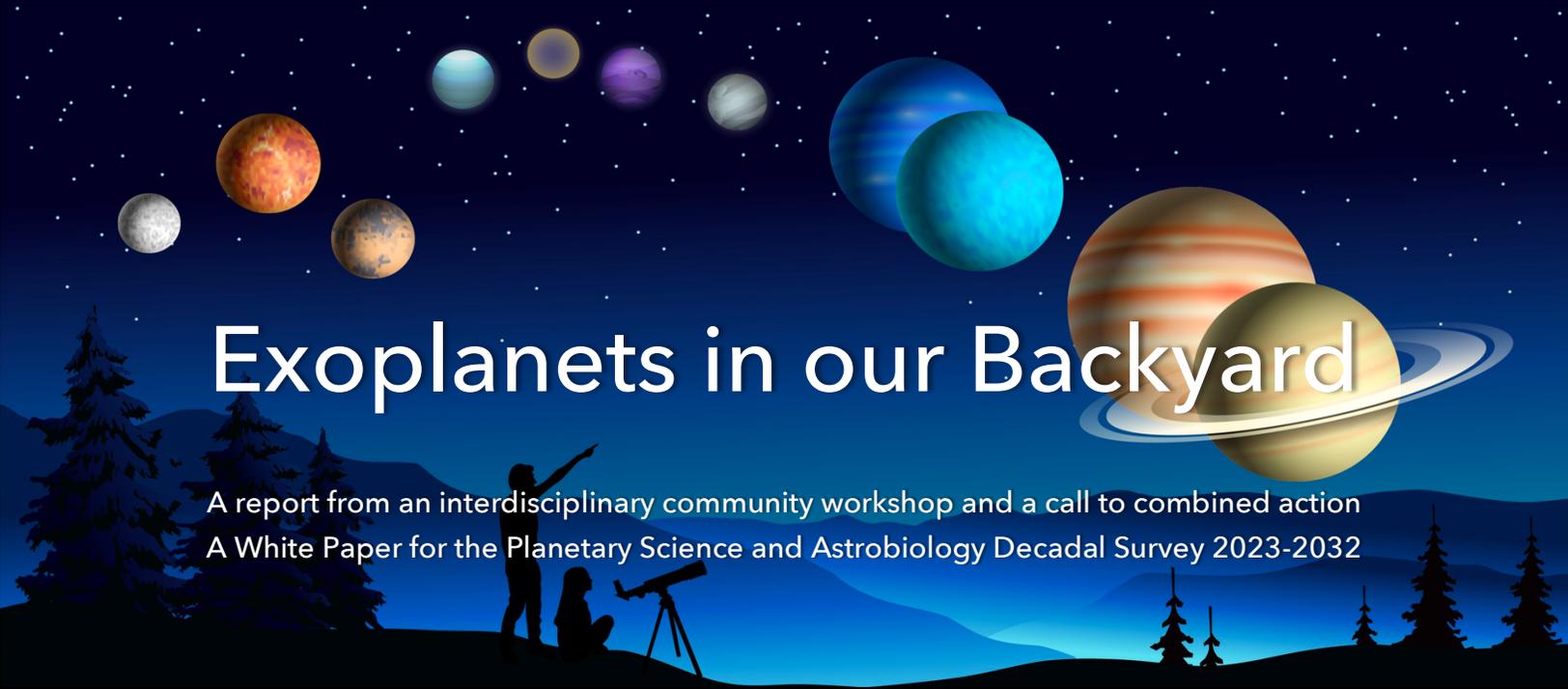

# Exoplanets in our Backyard

A report from an interdisciplinary community workshop and a call to combined action
A White Paper for the Planetary Science and Astrobiology Decadal Survey 2023-2032

**Giada N. Arney** *(giada.n.arney@nasa.gov  301-614-6627 Goddard Space Flight Center (GSFC), Greenbelt, MD, USA)*

**Noam R. Izenberg** *(noam.izenberg@jhuapl.edu, 443-778-7918, Johns Hopkins University Applied Physics Laboratory (JHUAPL), Laurel, Maryland, USA. )*

**Co-authors:**

**Stephen R. Kane,** *University of California, Riverside, CA*
**Kathleen E. Mandt,** *JHUAPL, Laurel, MD*
**Victoria S. Meadows,** *University of Washington, Seattle, WA*
**Abigail M. Rymer,** *JHUAPL, Laurel, MD*
**Lynnae C. Quick,** *GSFC, Greenbelt, MD*
**Paul K. Byrne,** *North Carolina State University, NC*

**Co-Signers:** Kevin Baines (JPL/Caltech), David Baker (Austin College), Kimberly Bott (UCR), Dave Brain (U. Colorado), Karalee K. Brugman (Arizona State University, currently at Carnegie Institution for Science), Ludmila Carone (MPIA, Heidelberg), Knicole Colón (NASA GSFC), Shawn Domagal-Goldman (NASA GSFC), Chuanfei Dong (Princeton University), Paul Dalba (UC Riverside), Ryan Felton (CUA), Dawn Gelino (Caltech/IPAC-NExScI), Scott Guzewich (NASA GSFC), Nader Haghighipour (IfA, Univ. Hawaii), Hilairy Hartnett (Arizona State University), Chao He (Johns Hopkins University), James W. Head (Brown University),  Michelle Hill (UC Riverside), Timothy Holt (Univ. Southern Queensland), Devanshu Jha (MVJCE), Der-you Kao (NASA GSFC/UMD), Walter S. Kiefer (Lunar and Planetary Institute/USRA), Scott King (Virginia Tech), Byeongkwan Ko (Arizona State University), Erika Kohler (NASA GSFC), Ravi Kopparapu (NASA GSFC), Jacob Lustig-Yaeger (University of Washington), Chuhong Mai (Arizona State University), Sarah Marcum (NASA GSFC/SURA), Mark Marley (NASA ARC), Emily Martin (UCSC), Laura Mayorga (Center for Astrophysics | Harvard & Smithsonian), Jonathan Mitchell (UCLA), Sarah E. Moran (JHU), Julianne I. Moses (Space Science Institute), Niki Parenteau (NASA ARC), Max Parks (NASA GSFC/UMBC), Daria Pidhorodetska (NASA GSFC/UMBC), Jani Radebaugh (BYU), Darin Ragozzine (BYU), Laura Schaefer (Stanford), Sang-Heon Dan Shim (Arizona State University), Kristin Showalter Sotzen (JHUAPL), Kathleen E. Vander Kaaden (JETS-NASA JSC)**,** Monica Vidaurri (Howard University/NASA GSFC)**,**  Eric T. Wolf (University of Colorado, Boulder)

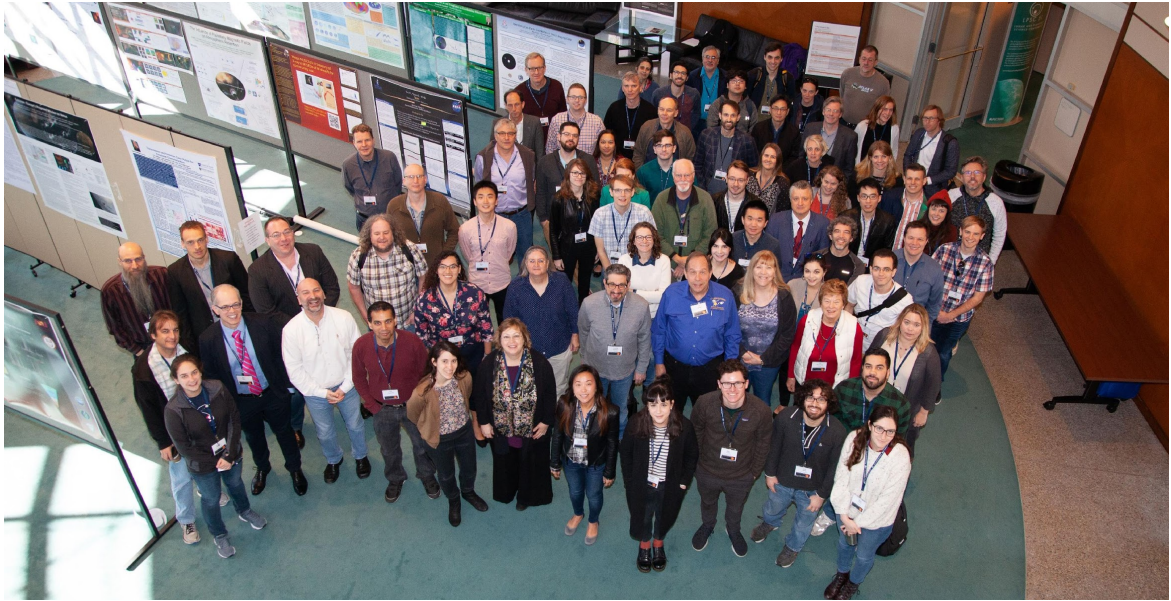

**Exoplanets in our Backyard**:
A report from an interdisciplinary community workshop and a call to combined action.

Key Points:
- **Both planetary and exoplanetary researchers benefit from interactions between the two communities.**
- **The Exoplanets in our Backyard Workshop was an exemplar interdisciplinary and interdivisional education, networking, and collaboration event.**
- **Key scientific and programmatic findings from the workshop show how NASA and the science community can encourage and nurture research at the intersection of the solar system and exoplanet fields.**

**Interdisciplinarity Connects Planets Near and Far**

There is a growing awareness of the power of systems- and process-based approaches to understanding planets, and it is increasingly clear that many of the most exciting and fruitful areas of research lie at the intersection of the traditionally separated domains of solar system and exoplanet research. Within our solar system, we find numerous examples of diverse worlds that can be studied close-up, showing us how planets operate on a level unthought of even a few decades ago. Comparative planetology between these worlds continues to empower a deeper understanding of the processes and phenomena that shape planets on wider scales than a single planet can tell us. Beyond our "cosmic backyard", more than four thousand exoplanets have been discovered as of July 2020. Despite the significant data limitations inherent to studying any individual exoplanet, these planets have the advantage of their vast numbers and can speak to planetary statistics and processes writ large. The same comparative planetology we apply within our solar system can be applied within and across these other planetary systems to reveal what is "rare" and what is "typical" on scales of individual worlds to the planetary system level—even showing us examples of worlds and processes unlike those seen in our solar system.

The Exoplanets in our Backyard meeting [1] was born out of a recognition of the value and potential of interdisciplinary, cross-divisional exoplanet and solar system research, and to encourage and grow the community of researchers working at this intersection. This first-ever inter-assessment group (AG) meeting (organized by members of the Venus Exploration, Outer Planets, and Exoplanet AGs, or VEXAG, OPAG, and ExoPAG, respectively), successfully brought together solar system and exoplanetary scientists from different backgrounds and NASA divisions, fostered communication between researchers whose paths had never crossed at a meeting before, and spurred new collaborations. The meeting was held at the Lunar and Planetary Institute in Houston, TX on February 5–8, 2020 immediately following the OPAG meeting hosted at the same location. The meeting was attended by approximately 110 scientists on site, and 20–30 online participants. The success of this meeting should be capitalized upon and its momentum carried forward to promote fruitful scientific and programmatic discussion, partnerships, and research going forward. This white paper summarizes the meeting, and discusses the findings and action items that resulted.

**Format of the Meeting**
The meeting opened with three overview talks on inner solar system planets, outer solar system planets, and exoplanets. These talks were focused to share material that might be relevant to other fields. The first morning also featured a NASA town hall with representative NASA leadership (Dr. Lori Glaze, Director of NASA's Science Mission Directorate Planetary Science Division, and Dr. Douglas Hudgins, NASA's Exoplanet Program's and the TESS mission's Program Scientist) discussing future NASA directions and fielding questions. To promote discourse on the interplay of planetary processes and phenomena rather than on singular planetary bodies, the bulk of the meeting was structured into six major scientific sessions on selected processes and themes rather than individual targets:
- Formation and Evolution of Planets
- Interior and Surface Processes
- Planetary Atmospheres Thick and Thin
- Star-Planet Interactions
- Habitability and Astrobiology Near and Far
- Missions

Each session featured four talks, after which all four speakers came to the stage for a mini-panel featuring further discussion and audience questions. This format, in contrast to the traditional talk–Q&A format, had the advantage of promoting extended dialog between panelists and the audience. The evening of the meeting's first day featured a lively public talk by Emily Lakdawalla in front of the giant IMAX screen at Space Center Houston to an audience of approximately 140 scientists, community members, and teachers from a concurrently co-located meeting. The second afternoon of the meeting featured a participant-led "Unconference" session—where participants met in groups to discuss crowd-sourced topics. These selected topics included "exomoons", "How important are exoplanet ages?", "ethics/policy/diversity," and "How can exoplanets help us better understand solar system planets and vice versa?". The second day concluded with a fruitful poster session preceded by minute-long flash talks. The last day of the meeting wrapped up with a final discussion/findings working lunch.

**Call to Action**

During the extended community discussions, attendees of the meeting identified several areas in which action is needed to ensure continued successful exoplanet/solar system research. Specifically:

**1. Cross-Divisional Cooperation is Vital**

The Exoplanets in our Backyard Workshop, as the first-ever inter-AG workshop, successfully brought together Solar System and exoplanetary scientists from different backgrounds and NASA divisions. The success of this workshop *and the strong cross-community support determined by post-conference survey* (results below), has resulted in a new cross-community Slack workspace[1] a special session at the 2020 AGU Fall Meeting, and a new ExoPAG Science Interest Group[2] jointly led by members of the exoplanet and solar system communities (this group is open to all members of the exoplanet, solar system, Earth science, and heliophysics communities). The momentum of Exoplanets in our Backyard should be carried forward and capitalized upon to promote science and programmatic discussion and partnerships, with planetary scientists informing exoplanetary research, modeling and observations, and exoplanetary scientists informing planetary research and missions. We point to the success of the NASA Astrobiology Institute, the Nexus for Exoplanet System Science (NExSS), and the Comparative Climates of Terrestrial Planets conferences to highlight the value of cross-divisional, interdisciplinary science. **The inter-AG, inter-division participation and cooperation of the community and support from NASA headquarters was essential to the meeting's success—and will continue to be so going forward for future similar meetings.**

**2. Enhancing Communication Between Our Communities**

Communication between the planetary and exoplanetary communities is essential to scientific progress in both disciplines and needs to be enabled, encouraged, and facilitated. Attendees at the workshop identified several areas of difficulty working across disciplines, including:
- Not knowing field terminology
- Not understanding key paradigms, ongoing puzzles, or the history of idea development in a given field
- Not understanding cultural aspects of a field that practitioners take for granted
- Not understanding limitations/assumptions in another discipline/community's models
- Not understanding where to find data from another discipline/community
- Not understanding limitations/relevant quality of another discipline/community's data

In response to this need for better and more opportunities for communication, the previously mentioned Slack workspace was set up in Feb. 2020 to provide an informal opportunity to interact with and learn from scientists in other disciplines and now includes 127 participants from multiple scientific communities. Work is still needed to recruit additional participants and develop cross-field tutorials, but this space has already proved useful for crafting this white paper, among others. Furher support could come from review papers that would provide a clear, jargon-free description understandable to someone outside the field, and targeted to focus on aspects another field would need to know to be able to interact more effectively.

---

[1] Contact the authors of this whitepaper to join.
[2] https://exoplanets.nasa.gov/exep/exopag/sigs/

Additional support is required for communications and documents regarding current states of knowledge in (exo)planetary science, including information on measurement limitations, guidance on available databases and reliability criteria, and methods to bridge/modify/improve planetary and exoplanetary models. In particular, there is interest in development of exoplanet databases that include detailed host star information. Existing cross-disciplinary outlets (e.g. NExSS) could assist in these efforts and serve as a resource hub for both the exoplanet and solar system communities. **NASA should encourage and support these communication efforts. Furthermore, to enhance communication across fields, we encourage the planetary science community to post their work to arXiv astro-ph.EP (Earth & Planetary Sciences).**

### 3. Inter-Community Participation in NASA Missions

The community represented in and by the Exoplanets in our Backyard workshop supports the development of opportunities for participation by exoplanet scientists in heliophysics, Earth and solar system exploration missions, as well as the corresponding participation of planetary scientists, Earth scientists and heliophysicists in exoplanet-relevant missions. These opportunities are currently rare or do not exist, but could include the routine addition of cross-discipline members to Science and Technology Development Teams for mission concept development, and as Participating Scientists on active missions. **NASA should enable and normalize cross-disciplinary participating scientist opportunities to provide additional insights into synergistic science for astrophysics and planetary exploration platforms, and to greatly enhance the overall science return from NASA missions.**

### 4. Growing the Interdisciplinary Early Career Community

Students and other early career researchers (postdocs, post bacs) were essential to the success of the meeting (representing 40% of submitted abstracts) and are vital to the future success and growth of the cross-disciplinary community fostered by Exoplanets in our Backyard. *Through generous funding from both NASA's Planetary Science Division and the Astrophysics Division's Exoplanets Exploration Office, Exoplanets in our Backyard was able to provide travel grants to 26 students and early career scientists.* Growing a community of scientists who learn to participate in interdisciplinary, cross-divisional research at early stages of their careers will help establish interdisciplinary research as the norm rather than the exception in the future. **Involvement of the student and early career community in future interdisciplinary meetings such as Exoplanets in our Backyard should be supported by NASA to foster the next generation of interdisciplinary scientists**.

This model has already been applied successfully in the astrobiology community, through venues including the interdisciplinary Astrobiology Graduate Conference (AbGradCon) [2-4]. AbGradCon is exclusively organized and attended by students and early career researchers with the dual goals of providing a space to share research among peers and growing a community of next-generation astrobiologists. **We also encourage the solar system and exoplanet science** communities to explore new interdisciplinary early career-only meetings to replicate the success of AbGradCon in fostering an emergent interdisciplinary community.

## 5. Growing a Diverse and Inclusive Community

There are challenges inherent to overcoming the barriers traditionally separating the exoplanet and solar system fields. While diverse suites of skill sets and perspectives are needed to overcome these challenges, equally important is the recognition that a just and welcoming scientific community is needed to overcome long standing systemic biases and exclusion. Naturally, the diversity that the community really needs not only encompasses different scientific backgrounds (e.g. astronomers, planetary scientists, geoscientists, heliophysicists, etc), but also other axes of diversity, including but not limited to gender, race, disability, and career stage. The Exoplanets in our Backyard organizing committee made a best effort to assemble a balanced and inclusive program reflective of the community. While gender can often be inferred from the names of invited and contributed presenters, we used google to ensure that our invited and contributed speakers represented a range of racial and ethnic backgrounds. However, we acknowledge that this method is imperfect, especially for nonbinary individuals [5], and we welcome input for how to improve. At the Exoplanets in our Backyard meeting, 50% of talks were given by inferred female speakers, and approx. 20% by inferred members of underrepresented minority groups. **We encourage all committees and scientific meetings to consider axes of diversity, especially the Planetary Decadal Survey. We also look to and support white papers that discuss strategies for improving diversity and inclusivity in our communities.**

## 6. Cross-Participation in Decadal Surveys

Cross-division participation in Decadal Surveys is extremely important for fields like exoplanet research, which is at the crossroads of all four Divisions represented by these surveys. The Astro2020 Decadal Survey included panel members from the planetary and heliophysics communities to provide critical input on advancing astrobiology and exoplanet research and to represent the interests of the planetary science community in ensuring that the management of Astrophysics resources, like HST and JWST, are responsive to the planetary community. **We strongly support and encourage continuing this cross-participation in future Astrophysical Decadal Surveys. Exoplanet experts should also serve on Planetary Science and Heliophysics Decadal Survey panels to provide input into the observations needed from missions led by these Divisions to help advance exoplanet science.**

## 7. Cross-Divisional Research Opportunities

The Exoplanets in our Backyard Workshop identified several avenues for ongoing research cooperation. **We encourage NASA cross-divisional R&A programs to take actions to address these research gaps:**

    **i.** Support and connect laboratory research with the observing communities (solar system and exoplanetary) through study of fundamentals and basic research (e.g., equations of state for $H_2O$, He, H, and mixtures; atmospheric dynamics experiments; thermal experiments; molecular opacities; optical constants; aerosol properties; etc.,).

    **ii.** Support comparative planetology research with modeling, field and data analysis efforts by improving the fidelity, complexity, and applicable parameter space of solar system and exoplanet models that simulate, e.g., atmosphere, magnetosphere, thermal, surface, and

interior processes and evolution; star/planet interaction; etc. Encourage cross-community comparison of models and data and development of models applicable to interpreting observable properties of planets (e.g. models that can be applied to highly irradiated planets on short orbits that will be observed with the James Webb Space Telescope in the near future).

**iii.** Support improved understanding of the characteristics of planet host stars, including our Sun, and star–planet interactions: e.g., stellar system metallicity, age, atmospheric escape processes, planetary and stellar magnetic fields, stellar wind and stellar storm activity, etc.

**iv.** Support cross-Divisional participating scientist and data analysis programs between Planetary Science, Heliophysics, Earth Science, and Astrophysics to advance space science research in the solar system and beyond.

**Success of the Meeting**

We are encouraged by the success of the Exoplanets in our Backyard meeting evidenced by the results of our post meeting survey (Figs. 1-3), and we strongly encourage the solar system and exoplanet communities to continue to establish venues for communication and collaboration. A follow-on meeting is being planned, and lead organizers have been identified.

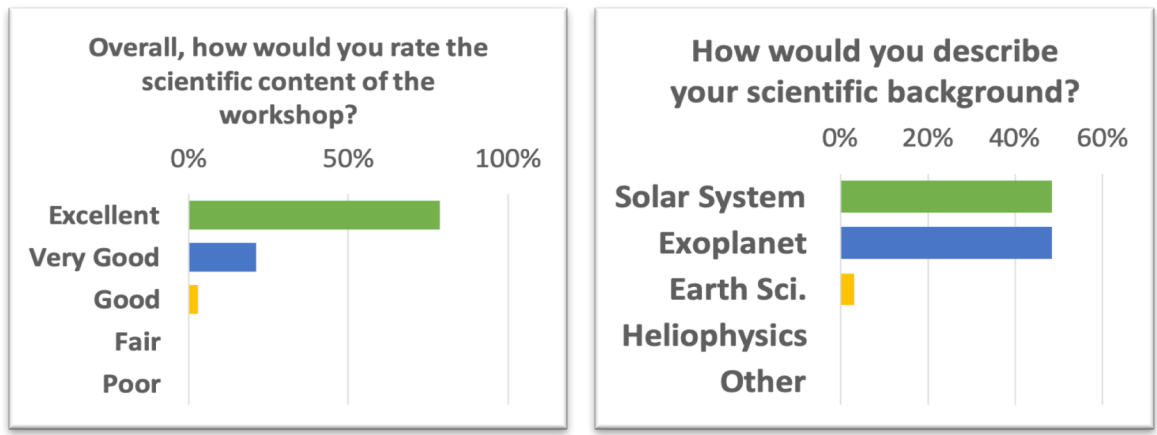

**Figure 1**. *(Left) The survey respondents rated the scientific content of the workshop favorably. (Right) Respondents were exactly split between the solar system and exoplanet communities, signaling that the meeting had successfully provided a bridge between these communities.*

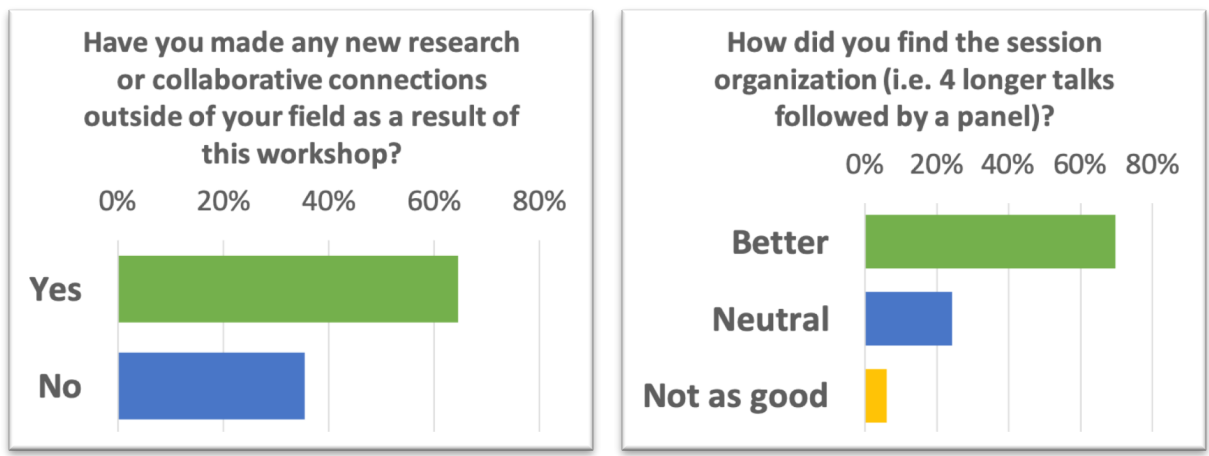

**Figure 2.** *(Left) More than half of respondents made a new collaboration or research connection outside of their field as a result of the meeting. (Right) The question and panel format for the science sessions was viewed favorably.*

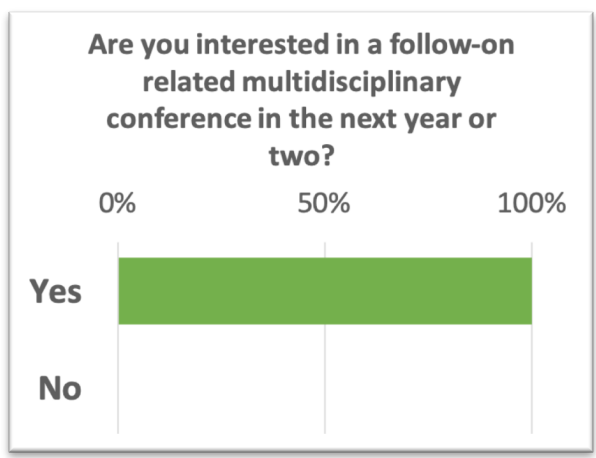

***Figure 3.*** *There is a strong desire for more meetings of this kind.*

**The Future**

Ultimately, exoplanet and solar system science must support each other as we move towards a more unified understanding of planets within our "backyard" and beyond. The Exoplanets in our Backyard meeting provided a successful bridge towards merging these communities, but this is just the beginning. The term "planetary science" is often taken as synonymous with solar system science, but we hope this definition will be broadened as exoplanet researchers are welcomed into the fold. The future of planetary science reaches from Earth to the edge of the cosmos; its power is manifested in linking the stunning detail of the worlds up close to the throng of billions of planets circling other stars. **After all, a planet is a planet no matter how far.**

**References**

**[1]** Exoplanets in our Backyard meeting website, with links to abstracts and recorded presentations. https://www.hou.usra.edu/meetings/exoplanets2020/
**[2]** AbGradCon   https://www.abgradcon.org/
**[3]** Som, S. M., Domagal-Goldman, S., Wright, K., Boldt, M., & Antonio, M. (2009). The Astrobiology Graduate Student Conference (AbGradCon). In *60th International Astronautical Congress 2009, IAC 2009* (pp. 8390-8397).
**[4]** McGonigle, J. M., Motamedi, S., & Rapf, R. J. (2019). Astrobiology Graduate Conference: A 15 Year Retrospective. *ACS Earth and Space Chemistry* 3 (12), 2675-2677
**[5]** Rasmussen, K. C. et al. (2019).  The Nonbinary Fraction: Looking Towards the Future of Gender Equity in Astronomy. https://arxiv.org/pdf/1907.04893.pdf